\documentclass{ws-mpla}

\begin{document}

\markboth{Yu-Xiao Liu,  Xin-Hui Zhang,  Yi-Shi Duan} {Detecting
Extra Dimension by Helium-like Ions}

\catchline{}{}{}{}{}

\title{Detecting Extra Dimension by Helium-like Ions}

\author{
    \footnotesize Yu-Xiao Liu \footnote{Email:
    liuyx@lzu.edu.cn},
    \footnotesize Xin-Hui Zhang \footnote{Corresponding author.
    Email: zhangxingh03@lzu.cn},
    \footnotesize Yi-Shi Duan}

\address{Institute of Theoretical Physics, Lanzhou University,
Lanzhou 730000, China}

\maketitle


\begin{abstract}

    Considering that gravitational force might deviate from Newton's
    inverse-square law and become much stronger in small scale, we
    present a method to detect the possible existence of extra
    dimensions in the ADD model. By making use of an effective
    variational wave function, we obtain the nonrelativistic ground
    energy of a helium atom and its isoelectronic sequence. Based on
    these results, we calculate gravity correction of the ADD model.
    Our calculation may provide a rough estimation about the magnitude of
    the corresponding frequencies which could be measured in later
    experiments.

\end{abstract}

\ccode{PACS Nos.: 04.80.-y; 04.50.+h; 31.15.Pf.\\}

\maketitle

There are at least two seemly functional energy scales in nature,
the electroweak scale $m_{EW}\sim{10^2}$ Gev and the Planck scale
$M_{pl}\sim{10^{19}}$ Gev, where gravity becomes as strong as the
gauge interactions. Over the last two decades, explaining the
smallness and radiative stability of the hierarchy
$m_{EW}/M_{pl}\sim{10^{-17}}$ has been one of the greatest driving
forces behind the construction of theories beyond the Standard Model
(SM). N. Arkani-Hamed, S. Dimopoulos and G. Dvali proposed an
exciting model (ADD model) \cite{ADD1,ADD2,ADD3}. 
In the ADD model, a very simple idea is to suppose there are $n$
extra compact spatial dimensions of radius $R$ and to assume that
$M_{pl(4+n)}$ is around the scale of $m_{EW}$. The weakness of
four dimensional gravitational force is explained as that the
origin strong high dimensional one leaks into extra dimensions. If
two test massed of mass $m_1$, $m_2$ placed within a distance
$r\ll R$, they will fell a gravitational potential dictated by
Gauss's law in (4+n) dimensions
\begin{equation}
V(r)\sim\frac{m_1m_2}{M_{pl(4+n)}^{n+2}r^{n+1}}.\ \ \ (r\ll R)
\end{equation}
On the other hand, if the masses are placed at distances $r\gg R$,
their gravitational flux lines can not continue to penetrate in
the extra dimensions, and the usual ${1}/{r}$ potential is
obtained
\begin{equation}
V(r)\sim\frac{m_1m_2}{M_{pl(4+n)}^{n+2}R^nr},\ \ \ (r\gg R)
\end{equation}
in which the radius of extra dimensions $R$ is determined by
$M_{pl(4+n)}$ as
\begin{equation}
R=\frac{1}{\pi}M^{-(1+\frac{2}{n})}_{pl(4+n)}G_4^{-\frac{1}{n}}.\ \
\ (c=1,\hbar=1)
\end{equation}
If $M_{pl(4+n)}\sim{1}$Tev, one can obtain $R \sim
\frac{1}{\pi}10^{-17+\frac{32}{n}}\textrm{cm}$. For $n=1$,
$R\sim10^{12}$ cm implying deviations from Newtonian gravity over
solar system distances, so this case is empirically excluded.
Particularly, the $n = 2$ case implies sub-millimeter extra
dimensions, while the experimental conditions for testing Newton's
inverse-square law (ISL) by torsion pendulum method was just about
to be available when ADD's proposal appeared. In fact, for all
$n\geq{2}$, however, the modification of gravity only becomes
noticeable at distances smaller that those currently probed by
experiments will be performed in the very near future. 
Therefore, many people have devoted to the search of deviation from
ISL as well as extra dimensions during the past few years
\cite{JCLong,CDHoyle}. Luo and Liu \cite{luo1,luo2} first proposed
the idea of detecting the deviation of Newton¡¯s inverse-square law
as well as exploring extra dimensions by spectroscopy experiments
and took hydrogen-like and muonic atomic system as illustrations.
Here we wonder if one can detect the extra dimension by
spectroscopic experiments due to a helium or helium-like ions.

From the early days of the quantum mechanics the ground state
ionization energy of a helium atom was a benchmark for approximate
methods of solving nonrelativistic Schr\"{o}dinger equation for a
few-body system. One of the earliest variational calculations has
been performed by Hylleraas \cite{Hyllerass} in 1929. In 1957,
Kinoshita \cite{Kinoshita} obtained higher order corrections
including the Lamb shift calculations confirmed a very good
agreement with the best experimental value. 
We would like to mention here the two most recent calculations. The
first is aimed to elaborate an efficient variational method for the
many electron atoms. The second is to find an effective and
economical way for studying a helium and helium-like two electron
atoms. Several techniques for example the finite-element method
\cite{JAckermann1998} and the hyperspherical harmonic method
\cite{RKrivec} have been used for such system, but variational
method is the most powerful tool for studying this problem and much
work has been carried out \cite{Pekeris,DEFreund,G.W.F
Drake,LiuHelium2004,LiuHelium2005}.

In this paper, we adopt a simple effective trial wave function and
calculate the variational ground state energy of a helium atom and
helium-like ions. Based on this trial wave function and making use
of the perturbation method, we obtain the gravity corrections of the
ADD model and the contribution to spectroscopy. In these
calculations, we use the Mathematic software, which is a tool of
symbolic calculation. All the results are exact except for that of
the wave function cutoff, which would bring an error but the error
can be controlled by setting the precision.

It is known that the non-relativistic Schr\"{o}dinger equation for
two-electron helium-like systems with nuclear charge $Z$ can be
written as ( in atomic units)
\begin{equation}
 H\psi(\mathbf{r}_1, \mathbf{r}_2)=E\psi(\mathbf{r}_1, \mathbf{r}_2),
\end{equation}
\begin{equation}
 H=\frac{1}{2}\mathbf{p}_1^2+\frac{1}{2}\mathbf{p}_2^2
   +\frac{1}{M}\mathbf{P}^2-\frac{Z}{r_1}
   -\frac{Z}{r_2}+\frac{1}{r_{12}},
\end{equation}
where $\mathbf{p}_{1,2}$ are the momenta of the electrons,
$\mathbf{P}=-\mathbf{p}_1-\mathbf{p}_2$ is the momentum of the
nucleus, $\mathbf{r}_1$ and $\mathbf{r}_2$ denote the positions of
the electrons with respect to the nucleus, and the finite
nucleus-to-electron mass ratio is given by $M\equiv
{m_{nucl}}/{me}$. To solve the above Schr\"{o}dinger equation for
the ground states, we employ variational approach by finding the
stationary solutions of the following energy functional
\begin{equation}
E=\textrm{min}\frac{\int\psi^*H\psi{d\tau}}{\int\psi^*\psi{d\tau}},
\label{variationalproblem}
\end{equation}
in which the volume element $d\tau=8\pi^2r_1r_2dr_1dr_2dr_{12}$. In
order to perform variational calculation, we need to make a
judicious choice for the wave function. To this end we adopt a
simply effective variational wave function, which contains  a
flexible scaling parameter $k$
\begin{eqnarray}
 &&\psi(ks,kt,ku)=e^{-ks}\sum{C_{lmn}}(ks)^l(kt)^{2m}(ku)^n,
 \label{wavefunction}\\
 &&s=r_1+r_2,\ \ \  t=-r_1+r_2,\ \ \ u=r_{12}, \label{rst}
\end{eqnarray}
where $s$, $t$ and $u$ are called the Hylleraas coordinates
\cite{Hyllerass} and the number of the terms of the polynomial in
$\psi$ is denoted as $N$. Under the Hylleraas coordinates, the
kinetic energy $K$,
 the potential energy $P$ and the denominator
 $W=\int\psi^*\psi{d\tau}$ in Eq. (\ref{variationalproblem})
can be expressed as follows, respectively
\begin{eqnarray}
 K&=&\int^\infty_{0}ds\int^s_0 du\int^u_{-u}dt\pi^2\{u(s^2-t^2)[(\partial_s\psi)^2
\nonumber \\
&&+(\partial_u\psi)^2+(\partial_t\psi)^2]-2s(t^2-u^2)\partial_s\psi\partial_u\psi \nonumber\\
&&+2t(s^2-u^2)\partial_t\psi\partial_u\psi+2u[(s^2-u^2)(\partial_s\psi)^2 \nonumber \\
&&-(t^2-u^2)(\partial_t\psi)^2]/M\},\label{K}\\
P&=&-\int^\infty_0ds\int^s_0du\int^u_{-u}dt\pi^2(4Zsu-s^2+t^2)\psi^2, \label{P}\\
W&=&\int^\infty_0ds\int^s_0du\int^u_{-u}dt\pi^2u(s^2-t^2)\psi^2.
\label{W}
\end{eqnarray}
Substituting Eqs. (\ref{wavefunction}), (\ref{rst}) into Eq.
(\ref{variationalproblem}) and using the variational method, we
calculate the ground state energy of a helium atom and helium-like
ions, and obtain the approximate wave function $\psi$. For
definition and simplicity, in the following calculation, we take
$N=10$.

The above work is performed by a program, which is written by
Mathematic Language. 
In what follows, based on the above calculating results, we
investigate the spectroscopy correction from the ADD gravitational
potential energy. Even if gravity would become stronger in small
scale, it is still very weak compared to electromagnetic force. So
it is convenient to treat the gravitational potential as a
perturbation to calculate the energy correction of a helium atom and
helium-like ions. The correction of the ground state energy is
written as
\begin{equation}
\Delta{E}=8\pi^2\int\psi^*\hat{V}(r_1,r_2,r_{12})\psi{r_1}
r_2dr_1dr_2dr_{12}.
\end{equation}
The gravitational potential is given by \cite{potential1}
\begin{eqnarray}
\hat{V}= \left\{
  \begin{array}{l}
    -\sum_{i=1}^{2}\frac{G_{(4+n)}m_e m_{nucl}}{r_i^{n+1}}
          -\frac{G_{(4+n)}m_em_e}{r_{12}^{n+1}},
          ~~~~~~~~~~~~~~~~~~~~~~~~~r_{1,2}\ll R \\
    -\sum_{i=1}^{2}\frac{G_4 m_e m_{nucl}}{r_i}(1+\alpha{e}^{-\frac{r_i}{\lambda}})
          -\frac{G_4m_em_e}{r_{12}}(1+\alpha{e}^{-\frac{r_{12}}{\lambda}}),
          ~~r_{1,2}~\sim ~R \\
    -\sum_{i=1}^{2}\frac{G_4 m_e m_{nucl}}{r_i}
          -\frac{G_4m_em_e}{r_{12}}.
          ~~~~~~~~~~~~~~~~~~~~~~~~~~~~~~~~~~r_{1,2}\gg R \\
  \end{array}
\right.
\end{eqnarray}
We consider the potential for the case of compactification on an
n-dimensional torus and assume that the space-time is (4 +
n)-dimensional, where the n extra dimensions are compactified on
circles with radius $R$. Then the $(4+n)$-dimensional Newton's
constant is expressed as \cite{potential2}
\begin{equation}\label{g4}
G_{(4+n)}=\frac{2v_nG_4}{s_n},
\end{equation}
where $v_n=(2\pi{R})^n$ is the volume of $n$ torus and
$s_n={2\pi^{\frac{n+1}{2}}}/{\Gamma(\frac{n+1}{2})}$ is its surface
area ($\Gamma(n)=\int^\infty_0e^{-x}x^{n-1}dx$). When $r\sim{R}$,
for the shape of extra dimensions mentioned above, the familiar
Yukawa potential is adopted $\alpha=8n/3$ and $\lambda=R$. One can
notice that the integral diverges as $r_{1,2}\rightarrow{0}$. For
$n\geq{2}$, considering the atomic nucleus is not pointlike, we
introduce a safe cutoff value $r_m$ with the atomic nucleus size for
the lower limit of the integral. A convenient expression of $r_m$ is
$r_m=r_0A^\frac{1}{3}$, where $A$ is the mass number of the atomic
nucleus and $r_0$ is of size $\sim{10^{-13}}$cm.

If we consider the n-dimensional sphere of radius $R$ as our
compactification manifold, the Newton constant is
\begin{equation}\label{g44}
G_{(4+n)}=2R^n{G_4}.
\end{equation}
For the case of large extra dimensions, let us compactify the $n$
dimensions $y_\alpha$ by making the periodic identification
$y_\alpha\sim{y_\alpha} + L$. Following exactly the same procedure
as Ref. \refcite{ADD3}, one can obtain
\begin{equation}\label{g444}
G_{(4+n)}=\frac{4\pi{V_n}G_4}{S_{(3+n)}},
\end{equation}
where $V_n=L^n=(2\pi{R})^n$ and
$S_D=2\pi^{\frac{D}{2}}/\Gamma(\frac{D}{2})$ is the surface area
of the unit sphere in $D$ spatial dimensions. From Eqs. (\ref{g4},
\ref{g44}, \ref{g444}), since $\frac{2v_n}{s_n}\sim 2R^n
\sim\frac{4\pi{V_n}}{S_{(3+n)}}$, we only take example for an
n-dimensional torus.

We apply the above wave function in Eq. (\ref{wavefunction})
to two-electron atomic and ionic systems. In Table 1, we quote the
values of the ground state energy for a helium atom and its
isoelectronic sequence. For the helium atom we also quote the values
of Hartree-Fock (HF) theory \cite{Fischer}, the 3-parameter
Charatzoulas-Knowleswave (CK) function \cite{Caratzoulas}, the
variational perturbation results of Pan \cite{Pan} and the
1078-parameter Pekeris wave function \cite{Pekeris2}. For other
ions, we compare our calculated values with the variational
perturbation results of Pan \cite{Pan} and Aashamar \cite{Aashamar}.
The relative errors compared with Pekeris and Aashamar are also
given in the last column.

\vskip 4mm

\textbf{Table 1.} Nonrelativistic ground energies of a helium atom
and helium-like ions calculated with $\psi$ in Eq.
(\ref{wavefunction}) in atomic units and their comparisons with
other references. For the helium atom, the relative error is
compared with the value of Pekeris. For the helium-like ions, the
relative errors are compared with that of Aashamar.
\begin{center}
\begin{tabular}{|l|c|c|c|}
\hline\hline
~System~ & ~Wave function~ & ~Ground energy~ & ~Error(\%)~ \\
\hline
         & $\psi$ & $-2.90313$ & \\
         & HF \cite{Fischer} & $-2.86168$ & \\
~~~He       & CK \cite{Caratzoulas} & $-2.89007$ & $0.0203$ \\
         & Pan \cite{Pan} & $-2.89122$ & \\
         & Pekeris \cite{Pekeris2} & $-2.90372$ & \\ \hline
         & $\psi$ & $-7.27906$ & \\
~~~Li$^+$   & Pan & $-7.26820$ & $0.0117$ \\
         & Aashamar \cite{Aashamar} & $-7.27991$ & \\ \hline
         & $\psi$ & $-13.65440$ & \\
~~~Be$^{2+}$ & Pan & $-13.64416$ & $0.0086$ \\
         & Aashamar & $-13.65557$ & \\ \hline
         & $\psi$ & $-22.02949$ & \\
~~~B$^{3+}$ & Pan & $-22.01973$ & $0.0067$ \\
         & Aashamar & $-22.03097$ & \\ \hline
         & $\psi$ & $-32.40438$ & \\
~~~C$^{4+}$ & Pan & $-32.39511$ & $0.0058$\\
         & Aashamar & $-32.40625$ & \\ \hline
         & $\psi$ & $-44.77929$ & \\
~~~N$^{5+}$ & Pan & $-44.77035$ & $0.0048$ \\
         & Aashamar & $-44.78145$ & \\ \hline
         & $\psi$ & $-59.15416$ & \\
~~~O$^{6+}$ & Pan & $-59.14554$ & $0.0041$ \\
         & Aashamar & $-59.15660$ & \\\hline
\end{tabular}
\end{center}
\vskip 4mm

Furthermore, by making use of the above wave function, we calculate
the frequencies of the spectrum corresponding to the ADD gravity
correction. For a helium atom, we get $\Delta\nu\sim{10^{-6}}$Hz
with $n=2$, and $\Delta\nu\sim{10^{-11}}$Hz with $n=3$. Although
these corrections are much larger than the one calculated from the
exact ISL ($\Delta\nu\sim{10^{-23}}$) for a helium atom, they are
still too small to be observed in current experiments. So we must
find some ``large" corrections that cannot be simply explained by
the inconsistency of the accurate experimental data and Standard
Model theoretical values. We can choose high-$Z$ helium-like systems
(such as Ca$^{18+}$ and Pb$^{80+}$ ) to do the similar calculation.
In Table 2, the frequencies of the spectrum for two-electron atom
and ions corresponding to the ADD gravity corrections are presented
in detail. However, we should point out that for Pb$^{80+}$, since
the corresponding Bohr's radius is smaller than that of a helium
atom, the effect of gravity from the nucleus is much more important.
Anyway, if we use the simple perturbation theory, the main
correction still comes from the Schr\"{o}dinger term. Because the
masses of Ca and Pb are representatively about 40 and 207 times of
the mass of proton, the corrections of the frequencies for
Ca$^{18+}$, Pb$^{80+}$ are much larger. For Ca$^{18+}$, we have
$\Delta\nu\sim{10^{-2}}$Hz with $n=2$, and
$\Delta\nu\sim{10^{-7}}$Hz with $n=3$; For Pb$^{80+}$,
$\Delta\nu\sim{10^{0}}$Hz with $n=2$ and $\Delta\nu\sim{10^{-5}}$Hz
with $n=3$, while the corresponding corrections in hydrogen atom are
$10^{-8}$Hz and $10^{-13}$Hz respectively. Also note that the exact
ISL, that is, the $n=0$ case gives a correction as small as
$\Delta\nu\sim{10^{-24}}$ Hz for Hydrogen atom and
$\Delta\nu\sim{10^{-23}}$ Hz for a helium atom.

\vskip 4mm
\textbf{Table 2.} The frequencies of the spectrum for two-electron
atom and ions corresponding to the ADD gravity corrections
\begin{center}
\begin{tabular}[t]{|l|l|l|}
  \hline\hline
  ~System~ & \begin{tabular}{c}  ~~~~$\Delta\nu$(Hz) \\  ~~~~($n=2$) \\  \end{tabular}
           & \begin{tabular}{c}  ~~~~$\Delta\nu$(Hz) \\  ~~~~($n=3$) \\  \end{tabular} \\
  \hline
  ~~~He ~&~ $-1.1 \times 10^{-6}$ ~&~ $-1.5 \times 10^{-11}$ \\\hline
  ~~~Li$^+$ ~&~ $-6.7 \times 10^{-6}$ ~&~ $-9.9 \times 10^{-11}$ \\\hline
  ~~~Be$^{2+}$ ~&~ $-2.1 \times 10^{-5}$ ~&~ $-3.2 \times 10^{-10}$ \\\hline
  ~~~B$^{3+}$ ~&~ $-5.0 \times 10^{-5}$ ~&~ $-7.9 \times 10^{-10}$ \\\hline
  ~~~C$^{4+}$ ~&~ $-9.5 \times 10^{-5}$ ~&~ $-1.5 \times 10^{-9}$~ \\\hline
  ~~~N$^{5+}$ ~&~ $-1.8 \times 10^{-4}$ ~&~ $-2.9 \times 10^{-9}$~ \\\hline
  ~~~O$^{6+}$ ~&~ $-3.0 \times 10^{-4}$ ~&~ $-5.0 \times 10^{-9}$~\\\hline
  ~~~F$^{7+}$ ~&~ $-4.9 \times 10^{-4}$ ~&~ $-8.5 \times 10^{-9}$~ \\\hline
  ~~~Ca$^{18+}$ ~&~ $-1.1 \times 10^{-2}$ ~&~ $-2.0 \times 10^{-7}$~ \\\hline
  ~~~Pb$^{80+}$ ~&~ $-3.1 \times 10^{0~}~$ ~&~ $-7.4 \times 10^{-5}$~ \\\hline
\end{tabular}
\end{center}
\vskip 6mm

We have referred to many references
\cite{value1,value2,value3,value4,value5,value6,value7,value8}
about experiment values of fine structure interval in helium or
helium-like ions, which collected in table 3. Comparing these
experiment values to Table 2, we find that even for Pb$^{80+}$,
the large shift can not yet been detected because of the precision
of the experiment. So, the ADD model can not yet be ruled out in
this way. We expect some experiments will be done in this way in
the future.

To summarize, we have formulated a calculational scheme for
detecting the extra dimension by using the helium and helium-like
systems. To perform the variational calculation we have used the
variational forms with a flexible parameter $k$ for two-electron
correlated wave function taking into account the motion of the
nucleus. The results obtained by us for the ground state is quite
accurate. Furthermore, based on the results,  we illustrate the
method numerically by calculating the magnitude of the
corresponding frequencies. For Pb$^{80+}$, the frequency derived
from the ADD gravity correction is $\Delta\nu\sim{10^{0}}$Hz with
$n=2$ and $\Delta\nu\sim{10^{-5}}$Hz with $n=3$. The corrections
might be used to indirectly detect the deviation of ISL down to
nanometer scale and to explore the possibility of two or three
extra dimensions in ADD's model, while current direct gravity
tests cannot break through micron scale and go beyond two extra
dimensions scenario.

\vskip 4mm

\textbf{Table 3.} Experimental fine structure intervals for $He$,
Li$^+$, Be$^{2+}$, B$^{3+}$, F$^{7+}$.
\begin{center}
\begin{tabular}{|l|c|c|c|}
\hline\hline
~System~ & ~Interval~ & ~Experiment~ & ~Reference~ \\
\hline
         & $\upsilon_{01}$ & $29616943.01(17)\ kHz$ & Ref. \refcite{value1}\\
         & $\upsilon_{12}$ & $2291161.13(30)\ kHz$ & \\
~~~He    & $\upsilon_{01}$ & $29616951.66\mp0.70\ kHz$ &  Ref. \refcite{value2}\\
         & $\upsilon_{12}$ & $2291175.59\mp0.51\ kHz$ & \\
         & $\upsilon_{01}$ & $29616950.9\mp0.9\ kHz$ & Ref. \refcite{value3}\\
\hline
         & $\upsilon_{01}$ & $155704.2161(30)\ MHz$ & Ref. \refcite{value4}\\
         & $\upsilon_{12}$ & $-62678.3382(27)\ MHz$ & \\
~~~Li$^+$  & $\upsilon_{01}$ & $155704.27(66)\ MHz$ & Ref. \refcite{value5}\\
         & $\upsilon_{12}$ & $-62678.41(66)\ MHz$ & \\
\hline
             & $\upsilon_{01}$ & $11.5576605(7)\ cm^{-1}$ & Ref. \refcite{value4}\\
             & $\upsilon_{12}$ & $-14.892209(1)\ cm^{-1}$ & \\
~~~Be$^{2+}$ & $\upsilon_{01}$ & $11.5586(5)\ cm^{-1}$ & Ref. \refcite{value6}\\
             & $\upsilon_{12}$ & $-14.8950(4)\ cm^{-1}$ & \\
\hline
             & $\upsilon_{01}$ & $16.197573(2)\ cm^{-1}$ & Ref. \refcite{value4}\\
             & $\upsilon_{12}$ & $-52.661199(4)\ cm^{-1}$ & \\
~~~B$^{3+}$  & $\upsilon_{01}$ & $16.203(18)\ cm^{-1}$ & Ref. \refcite{value7}\\
             & $\upsilon_{12}$ & $-52.660(16)\ cm^{-1}$ & \\
\hline
             & $\upsilon_{01}$ & $-151.2466(1)\ cm^{-1}$ & Ref. \refcite{value4}\\
~~~F$^{7+}$  & $\upsilon_{12}$ & $-957.8487(2)\ cm^{-1}$ & \\
             & $\upsilon_{12}$ & $-957.883(19)\ cm^{-1}$ & Ref. \refcite{value8}\\
\hline
\end{tabular}
\end{center}
\vskip 4mm

This work was supported by the National Natural Science Foundation
of the People's Republic of China (No. 502-041016, No. 10475034 and
No. 10705013) and the Fundamental Research Found for Physics and
Mathematic of Lanzhou University (No. Lzu07002).

\end{document}